\documentclass[twocolumn,aps,showpacs,pra]{revtex4}
\usepackage{epsfig}
\usepackage[english]{babel}
\usepackage{latexsym}
\usepackage{graphics}
\usepackage{subfigure}
\usepackage{epsfig}
\usepackage{graphicx}
\usepackage{dcolumn}
\usepackage{amsmath}

\begin{document}
\title{High-Order Harmonic Generation and  Molecular Orbital Tomography:
Characteristics of Molecular  Recollision Electronic Wave Packets}
\author{Y. Chen$^{1,2}$, Y. Li$^3$ , S. Yang$^3$ and J. Liu$^{2,4}$}

\date{\today}

\begin{abstract}
We investigate the  orientation dependence of molecular high-order
harmonic generation (HHG)  both numerically and analytically. We
show that the  molecular recollision electronic wave packets (REWPs)
in the HHG are closely related to the ionization potential as well
as the particular orbital from which it ionized. As a result, the
spectral amplitude of the molecular REWP can be significantly
different from its reference atom (i.e., with the same ionization
potential as the molecule under study) in some energy regions due to
the interference between  the atomic cores of the  molecules. This
finding is important for  molecular orbital tomography using
HHG[Nature \textbf{432}, 867(2004)].

\end{abstract}
\affiliation{1.Graduate School, China Academy of Engineering
Physics, P.O. Box 8009-30, Beijing, 100088, P. R.
China\\2.Institute of Applied Physics and Computational
Mathematics, P.O.Box 100088, Beijing, P. R. China\\
3. Department of Physics, Hebei Normal University, Shijiazhuang,
050016, P. R. China\\4. Center for Applied Physics and Technology,
Peking University, 100084, Beijing, P. R. China} \pacs{42.65.Ky,
32.80.Rm} \maketitle

The orientation dependence of molecular  high-order harmonic
generation (HHG) in strong laser fields constantly attracts
attention in both experiment and theory\cite{Velotta,J. Itatani,X.
X. Zhou,C. B. Madsen,lein1,Bandrauk,Tsuneto,vozzi}. A new surge
was brought on by the recent emergence of its important
application in molecular orbital tomography. Itatani $et\ al$
showed that through calibrating the recollision electronic wave
packet (REWP) by a reference atom, the orientation dependence can
be used to reconstruct the shape of the highest occupied molecular
orbital (HOMO)\cite{ilzn04}. The molecular orbital tomography  is
based on the assumptions  that, i) a reference atom exists that is
mainly characterized by an ionization potential close to that of
the molecule and, ii) the spectral amplitude of the REWP for the
reference atom is always identical to that of the molecule,
independent of the alignment of the molecule\cite{ilzn04,S.
Patchkovskii,R. Torres,Le,S. Patchkovskii2}. Investigation of the
foundations of molecular orbital tomography  has led to valuable
insights into the mechanism of atomic and molecular HHG. For
example, Levesque et al.\cite{Levesque} conjectures a universality
in the behavior of atomic HHG spectra scaled by the recombination
transition atomic dipole moment. This  conjecture and many other
assumptions connected to the idea of molecular orbital tomography
call for a thorough theoretical examination.

In the present paper, we directly examine   the  second assumption
by numerically calculating the spectral amplitudes  for 1D and 2D
H$_2^+$ molecular ions with different internuclear distances and
laser intensities. Through  comparison with reference atoms that
share the same ionization potential, we find that the REWP of
H$_2^+$ is strongly dependent on its orientation in certain energy
regimes where the interference between the two atomic cores of the
molecules dominates\cite{Muth-Bohm}. Analytically, we apply the
Lewenstein theory to N$_2$ and CO$_2$ molecules that have different
HOMOs and internuclear distances. Our theoretical calculations
validate the assumption for N$_2$ in the plateau region of the
harmonic spectra that are of most crucial for tomographic
reconstruction of the orbital. But, our calculations invalidate the
assumption for CO$_2$, showing a strong orientation dependence of
the spectra in the plateau region due to the interference. This
finding suggests that for  molecules such as CO$_2$, to accurately
reconstruct its orbital using the HHG, the interference effects need
to be considered.

The Hamiltonian of  H$_2^+$ or hydrogen-like atoms studied here is
$ H(t)=\mathbf{p}^2/2+V(\mathbf{r})-\mathbf{r}\cdot \mathbf{E}(t),
$ with soft-Coulomb potential
$V(x)$$=$$\frac{-Z}{\sqrt{1.44+({x+R/2})^{2}}}+\frac{-Z}{\sqrt{1.44+({x-R/2})^{2}}}$
and $ V(x,y)$$=$$\frac{-Z}{\sqrt{0.5+({x+R/2})^{2}+y^2}}+
\frac{-Z}{\sqrt{0.5+({x-R/2})^{2}+y^2}}$ for the 1D and 2D case,
respectively, where $Z$ is the effective charge, $R$ is the
internuclear separation for H$_2^+$, and $R=0$ a.u. corresponds to
the hydrogen-like atom, (for $Z=1$ and  $R=2$ a.u., the ground
state energy for H$_2^+$ reproduced here is $E_{0}=1.11$ a.u. in
both cases). $\mathbf{E}(t)$ is the external electric field. Here,
we suppose that the molecular axis is coincident with the x-axis
and the external field is linearly polarized with an orientation
angle of $\theta$ to the molecular axis. In the following
calculations, the atom units of $\hbar=e=m_e=1$ are adopted. Our
calculation will be performed for $780$ nm trapezoidally shaped
laser pulses with a total duration of $10$ optical cycles and
linear ramps of three optical cycles. Numerically, the above
Schr\"odinger equation is solved by the operator-splitting method.
The fast Fourier transform (FFT) is employed to transform the wave
function between the position representation and the momentum
representation. The boundary is set as $\pm 200$ a.u. and the grid
number is $4096$. The ground state is obtained by propagation in
imaginary time. The numerical convergence is checked with using a
finer grid. The coherent part of the harmonic spectrum is obtained
from the Fourier transformed dipole acceleration expectation
value, and only the harmonics polarized parallel to the incoming
field are considered\cite{lein1}.

According to Ref.\cite{ilzn04}, the spectral amplitude $a(E_p)$ is
\begin{equation}
|a(E_p)|=\sqrt{S(\Omega)}/2\pi\Omega^2|\mathbf{d}|,
\end{equation}
where $\mathbf{d}(\mathbf{p})=\langle\mathbf{p}|\mathbf{r}|0\rangle$
is the transition dipole matrix element of the atom or molecule
between the ground state $|0\rangle$ and the continuum state
$|\mathbf{p}\rangle$, $S(\Omega)$ is the radiated harmonic signal,
$\Omega$ is the photon energy, and $E_p=\mathbf{p}^2/2$ is the
electronic kinetic energy with $\Omega=E_p+I_p$. $I_p$ is the
ionization potential of the atom or molecule.
\begin{figure}[t]
\begin{center}
\rotatebox{0}{\resizebox *{7.0cm}{3.0cm} {\includegraphics
{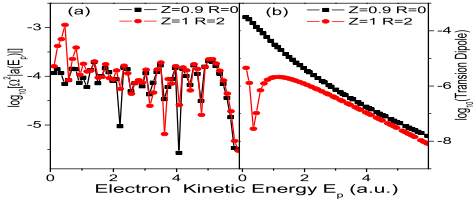}}}
\end{center}
\caption{(Color online) The spectral amplitudes
$\Omega^2|a(E_p)|=\sqrt{S(\Omega)}/|\mathbf{d}|$ (a) and transition
dipoles $|\langle0|x|\mathbf{p}\rangle|^2$(b) of $1D$ H$_2^+$ (the
red lines) and its reference atom (the black lines) with the
continuum $|\mathbf{p}\rangle$ obtained by the numerical method.}
\label{fig.1}
\end{figure}

Fig. 1(a) plots the spectral amplitudes $|a(E_p)|$ of 1D H$_2^+$
with $Z$$=$$1,R$$=$$2$ a.u. and its reference atom with $Z$$=$$0.9,
R$$=$$0$ a.u., where $|a(E_p)|$ is calculated  from the numerical
spectrum data by dividing by the transition matrix elements
$|\mathbf{d}|=|\langle0|x|\mathbf{p}\rangle|$ with the
$|\mathbf{p}\rangle$ obtained by  numerically diagonalizing the
field-free Hamiltonian $H_{0}=\mathbf{p}^2/2+V(x)$.  The
corresponding transition dipoles used in Fig. 1(a) are shown in Fig.
1(b) and will be discussed in detail bellow. The amplitude for
H$_2^+$ in Fig. 1(a) is  consistent with that of its reference atom
in the high energy region but different in the low energy region,
because it  shows a  pronounced peak structure, i.e., the amplitude
of the peaks is one-order of magnitude higher.

We now extend these calculations to 2D. Fig. 2 plots the spectral
amplitudes of 2D models of H$_2^+$ for varied molecular parameters
and field intensities. The spectral amplitudes $|a(E_p)|$ are
calculated from the numerical spectrum data by dividing by the
transition matrix elements in the length gauge
$|\mathbf{d}_{len}|=|\langle0|\mathbf{r}|\mathbf{p}\rangle|$ with
$|0\rangle$ obtained by the LCAO-MO approximation, and
$|\mathbf{p}\rangle$ obtained by the plane wave approximation and
the dispersion relation $\Omega=\mathbf{p}^{2}/2$\cite{Levesque}.
For $\theta=90^0$, the spectral amplitudes of the models of
H$_2^+$ and their corresponding reference atoms, which have been
shifted vertically to compensate for differing the overall
efficiency, are analogous (see  Fig. 2(a) and (b)). However, for
other angles, the most remarkable feature exhibited in Fig.
2(c)-2(f) is the peak structure  at the electron kinetic energy
that shifts rightward as the orientation angle increases. Except
for these peaks, the spectral amplitudes of the other energy
regimes are analogous. In addition, for the same  orientation
angle but different internuclear distance, the location of the
peaks is shifted. For example, at $\theta=50^0$, the peak is at
$E_p=1.85$ a.u. in Fig. 2(c), but it shifts to $E_p=3$ a.u. in
Fig. 2(d). For the  same internuclear distance,  at a fixed
orientation angle the location of the peaks is almost unchanged
when laser intensity varies, as revealed by Fig. 2(c) with
$I=8.5\times10^{14} W/cm^{2}$ and 2(e) with $I=5\times10^{14}
W/cm^{2}$. These calculations on 1D and 2D H$_2^+$ for diverse
internuclear distance and laser intensities show that the spectral
amplitudes of the molecules  are usually dependent on the
molecular orientation.

\begin{figure}[t]
\begin{center}
\rotatebox{0}{\resizebox *{8.5cm}{7.0cm} {\includegraphics
{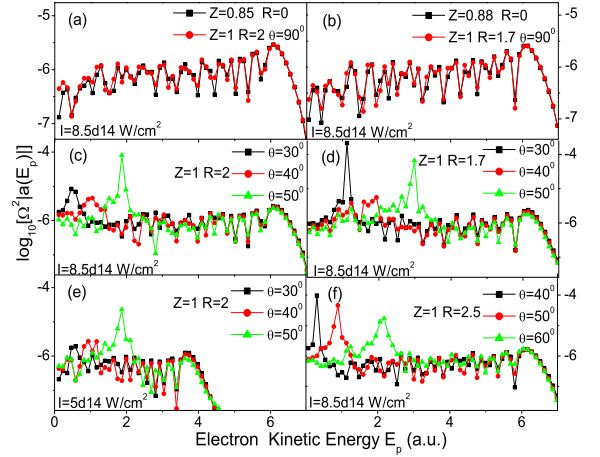}}}
\end{center}
\caption{(Color online) The spectral amplitudes
$\Omega^2|a(E_p)|=\sqrt{S(\Omega)}/|\mathbf{d}_{len}|$  of 2D
H$_2^+$ for varied molecular parameters and  field intensities
(indicated in each subpanel) calculated in the length gauge with the
dispersion relation $\Omega=\mathbf{p}^{2}/2$. In Fig. 2(a) and
2(b), we also plot the $|a(E_p)|$ of those corresponding reference
atoms for comparison.} \label{fig.2}
\end{figure}

Next, we explain the mechanism behind these phenomena according to
the Lewenstein model\cite{Lewenstein}, where the time-dependent wave
functions can be expanded  under the strong-field approximation,
$
|\psi(t)\rangle=e^{iI_pt}[a(t)|0\rangle+\int
d\mathbf{p}c_{\mathbf{p}}(t)|\mathbf{p}\rangle],
$
where $a(t)\approx1$ is the ground-state amplitude,  and
$c_{\mathbf{p}}(t)$ are the amplitudes of the corresponding
continuum states, and can be written in the closed form
\begin{eqnarray}
\begin{split}
c_{\mathbf{p}}(t)=i\int_0^t
dt'\mathbf{E}(t')\cdot\mathbf{d}(\mathbf{p}+\mathbf{A}(t)-\mathbf{A}(t'))\\
\times e^{-i\int_{t'}^{t}
[(\mathbf{p}+\mathbf{A}(t)-\mathbf{A}(t''))^{2}/2+I_p]dt''},
\end{split}
\end{eqnarray}
where $\mathbf{A}(t)$ is the vector potential of the laser field
$\mathbf{E}(t)$. The time dependent dipole moment can be written
as $
\mathbf{D}(t)=\langle\psi_{0}(\mathbf{r},t)|\mathbf{r}|\psi_{c}(\mathbf{r},t)\rangle,
$ where $\psi_{0}(\mathbf{r},t)$=$e^{iI_pt}|0\rangle$ denotes the
initial electronic state and
$\psi_{c}(\mathbf{r},t)$=$e^{iI_pt}\int
d\mathbf{p}c_{\mathbf{p}}(t)|\mathbf{p}\rangle$ denotes the REWP.
Eq. 2 shows that the time-dependent continuum amplitude
$c_{\mathbf{p}}(t)$ is not only dependent on the particular
ionization potential $I_p$  in  the exponent of the integrand, but
also depends on the ground state wave function through the
transition dipole $\mathbf{d}(\mathbf{p})$ in the prefactor of the
integrand. Thus,  the continuum amplitudes of the molecule and its
reference atom sharing the same ionization potential $I_p$ should
have the same time-dependent phase, but their transition dipoles
$\mathbf{d}(\mathbf{p})$ differ.

\begin{figure}[t]
\begin{center}
\rotatebox{0}{\resizebox *{7.5cm}{3.8cm} {\includegraphics
{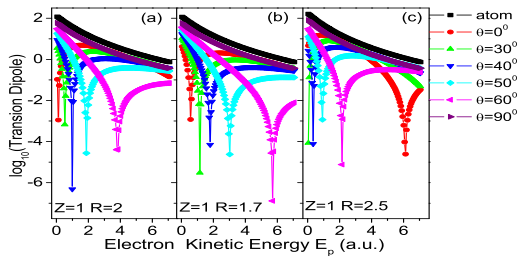}}}
\end{center}
\caption{(Color online) The transition dipoles
$|\langle0|\mathbf{r}|\mathbf{p}\rangle|^2$ along the laser
polarization direction of 2D H$_2^+$ with different $R$ and
orientation angles $\theta$ and their corresponding reference atoms,
calculated in the length gauge with the dispersion relation
$\Omega=\mathbf{p}^{2}/2$.} \label{fig.3}
\end{figure}

The above theoretical analysis has been checked by numerical
simulation. For the 1D case, Fig. 1(b)  clearly demonstrates that
the transition dipole for H$_2^+$  is different from that for its
reference atom. The transition dipole for H$_2^+$  shows a deep
hollow (the amplitude of the hollow may be up to four order of
magnitude lower) around $E_p=0.33$ a.u., but is analogous with its
reference atom in a broad higher energy region
($E_p=1.5\rightarrow 6$ a.u.). The ionization potential of H$_2^+$
is similar to its reference atom, and their transition dipoles
also are analogous in a broad high energy region. Therefor, we
expect that their spectral amplitudes $a(E_p)$, which are
determined by $c_{\mathbf{p}}(t)$, also should be analogous in the
high energy region, but different in the low energy region, as
shown by Fig. 1(a). In addition, the location of the hollow in
Fig. 1(b) corresponds to that of the peak in Fig. 1(a).

Fig. 3 shows the 2D transition dipoles used to produce Fig. 2. The
transition dipoles for the reference atoms (the black curves in
each subpanel of Fig. 3) are analogous, but the transition dipoles
for the models of H$_2^+$ with different $R$ show strong
orientation dependence. The deep hollow  moves as the angle
$\theta$ changes. But, at $\theta=90^0$, the transition dipoles
for the models of H$_2^+$ are similar to that of the  reference
atoms in the  energy region  $E_p$=$0\rightarrow 7$ a.u.. Thus we
expect that, for $\theta=90^0$, the spectral amplitude of H$_2^+$
should be identical with its reference atom, but for $\theta \ne
90^0$, they will show a large difference  in the energy region
where the deep hollow in the transition dipole of H$_2^+$ appears.
This theory is verified by the plot in Fig. 2.

The 1D and 2D numerical results show that intrinsic properties of
the molecules already enter the REWP. The deep hollows in the
transition dipoles of  H$_2^+$ with different $\theta$ can be read
from the interference effect as predicted by Muth-Bohm $et\
al$\cite{Muth-Bohm}. Under the LCAO-MO approximation, the ground
state of the diatomic molecule is $
|0\rangle=\sum_ia_{i}|\mathbf{\phi}_{0}(\mathbf{r}-\mathbf{R}/2)\rangle+b_{i}|\mathbf{\phi}_{0}(\mathbf{r}-\mathbf{R}/2)\rangle,
$ where $|\mathbf{\phi}_{0}(\mathbf{r})\rangle$ is the atomic
ground state and $a_{i}$ and $b_{i}$ are the normalization factors
with $a_{i}=b_{i}$ for symmetric mixing and $a_{i}=-b_{i}$ for
antisymmetric mixing. For symmetric mixing, e.g., H$_2^+$, the
transition dipole can be written as
\begin{eqnarray}
\mathbf{d}(\mathbf{p})=2\cos(\mathbf{p}\cdot\mathbf{R}/2)\sum_{i}a_{i}\mathbf{s}(\mathbf{p})+
i\mathbf{R}\sin(\mathbf{p}\cdot\mathbf{R}/2)\sum_{i}a_{i}\tilde{\mathbf{\phi}}_{0}.
\end{eqnarray}
Here $
\mathbf{s}(\mathbf{p})=\int[e^{-i\mathbf{p}\cdot\mathbf{r}}\mathbf{r}{\mathbf{\phi}}_{0}(\mathbf{r})]d\mathbf{r},
\tilde{\mathbf{\phi}}_{0}(\mathbf{p})=\int[e^{-i\mathbf{p}\cdot\mathbf{r}}{\mathbf{\phi}}_{0}(\mathbf{r})]d\mathbf{r}.
$ The  $\cos(\mathbf{{p}}\cdot \mathbf{{R}}/2)$ of the first term in
Eq. 3 represents the interference between the two cores of the
molecules\cite{Muth-Bohm}, and the second term is proportional to
the internuclear distance $R$ in Eq. 3, leading to the breakdown of
translation invariance. In our calculations, the second term is
omitted according to Ref.\cite{cj,Milosevic,W. Becker}. Thus, the
transition dipole of the molecule $\mathbf{d}(\mathbf{p})$ is
different from that of the reference atom $\mathbf{s}(\mathbf{p})$
due to the interference effect  from the term
$\cos(\mathbf{{p}}\cdot \mathbf{{R}}/2)$. The interference that
manifests itself as the hollows in the spectra of the transition
dipole or the peaks in the HHG spectral amplitudes for the
molecules, depends on the alignment angle $\theta$ and can be
located by $pR\cos(\theta)=(2n+1)\pi$ where $n$ is an integer. For
$R=2$ a.u. and $\theta=0^0$, the formula with the dispersion
relation $\Omega=\mathbf{p}^{2}/2$ predicts the hollow location of
$E_p=0.12$ a.u., which is close  to our numerical result of
$E_p=0.35$ a.u. in Fig. 1(b). If  the dispersion relation
$\Omega=\mathbf{p}^{2}/2+I_p$\cite{Levesque} is adopted, the
location is estimated at $E_p=1.23$ a.u., far away from the
numerical result. We have calculated the dependence of the minima
positions (in energy $E_p$) on molecular internuclear distance $R$
in the 1D case. It shows that for the larger $R$ (i.e., $R\geq3$
a.u). and larger momenta $p$ (i.e., $p\geq3$ a.u.), where the
LCAO-MO approximation for the ground state $|0\rangle$ and the plane
wave approximation for the continuum $|\mathbf{p}\rangle$ are
expected to be more applicable, the above simple  formula gives a
good estimation on the location of the minima, i.e., the deviation
is less than $1/10$.

In addition,  the transition dipoles for the reference atoms with
different ionization potentials, as shown in  Fig. 3(a) and 3(b),
are almost identical. We attribute the difference in  their  HHG
spectral amplitudes revealed  by comparing  Fig. 2(a) and 2(b)  to
the difference in  their ionization potentials. The dependence of
the spectral amplitude $a(E_p)$ on the ionization potential is
also evident in the current theories of tunnel ionization
(ADK\cite{Ammosov}, Yudin and Ivanov\cite{Gennady},
MO-ADK\cite{Tong}), where the ionization potential explicitly
appears in the exponent of the formula determining the cycle
dependence of the ionization rate. Thus, we  expect that the time
dependence of the continuum wave packet also relies  on the
ionization potential. Moreover, the present experiment for the
orientation dependence of ionization rate of CO$_2$ shows a
pronounced deviation  from the prediction by the MO-ADK theory.
The interference caused by the separated cores is not well
described by the simple theory\cite{Domagoj}. Thus, we expect that
the dependence of the continuum wave packet on the HOMO is also
not well described by the model. Based on this analysis, we
conclude that due to the interference effect, the ionization and
acceleration processes in the three-step model\cite{Corkum} also
play an important role in HHG from molecules. In fact, in our
calculations, the hollows in Fig. 3 for molecules are responsible
for the interference minima  in harmonic spectra predicted by Lein
$et\ al$\cite{lein1}.
\begin{figure}[t]
\begin{center}
\rotatebox{0}{\resizebox *{8.0cm}{6.0cm} {\includegraphics
{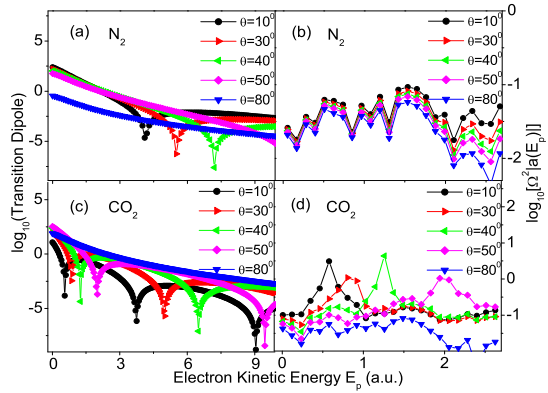}}}
\end{center}
\caption{(Color online) The transition dipoles
$|\langle0|\mathbf{r}|\mathbf{p}\rangle|^2$ along the laser
polarization direction ((a) and (c)) and spectral amplitudes
$\Omega^2|a(E_p)|=\sqrt{S(\Omega)}/|\mathbf{d}_{len}|$ ((b) and (d))
of  N$_2$ and CO$_2$ with different orientation angles $\theta$
calculated in the length gauge with $\Omega=\mathbf{p}^{2}/2$. The
laser intensity is $I=2\times10^{14} W/cm^{2}$ and wave length
$\lambda=800$ nm. The orbitals of N$_2$ ($2p\sigma_{g}$) and CO$_2$
($2p\pi_{g}$) are expressed in the LCAO-MO approximation.}
\label{fig.4}
\end{figure}

This analysis demonstrates that the spectral amplitude of the
molecular REWP is different from its reference atom due to the
interference. However, why can the molecular orbital tomography
experiment  reconstruct the HOMO of N$_2$ by its reference atom Ar
\cite{ilzn04}? In Fig. 4, we plot the transition dipoles
$|\mathbf{d}_{len}|^{2}=|\langle0|\mathbf{r}|\mathbf{p}\rangle|^2$
along the laser polarization direction ((a) and (c)) and the
spectral amplitudes $|a(E_p)|$ ((b) and (d)) of N$_2$ and CO$_2$
with different orientation angles $\theta$. The corresponding
harmonic spectra are obtained by the Lewenstein model with the
dispersion relation
$\Omega=\mathbf{p}^{2}/2$\cite{ilzn04,Levesque}. For N$_2$ with
$R=2.079$ a.u., the interference hollows  at different angles
$\theta$ should all appear in the high energy region, i.e., far
away from the plateau regime. The curves of the transition dipoles
in Fig. 4(a) are analogous to each other in the low energy region
($E_p\leq 2$ a.u.). Accordingly, the spectral amplitudes in Fig.
4(b) also are analogous to each other in the same energy region
within a vertical scaling factor. However, for CO$_2$ with
$R=4.38$ a.u., the curves of the transition dipoles  in Fig. 4(c)
show a large difference in the low energy region. The curves of
the spectral amplitudes in Fig. 4(d) also are obviously different,
showing a  sharp peak shifting toward the higher energy region as
the angle $\theta$ increases. We thus expect that the spectral
amplitude for N$_2$ in the low energy region is largely
independent of molecular orientation \cite{ilzn04}. This
conclusion is not applicable to CO$_2$, for which the interference
hollows for small orientation angles $\theta$ should appear in the
lower energy region corresponding to lower harmonic orders in the
plateau region\cite{Tsuneto,vozzi}. The spectral amplitude of
CO$_2$ in that energy region is largely related to its
orientation. Therefore, for some molecules, e.g., CO$_2$, the
interference effect should be included in  the orbital tomography
experiment to obtain an accurate HOMO wave function.

In summary, from our 1D and 2D numerical calculation and
analytical deduction, we find that the assumption of the
tomography method, i.e., that the recolliding electron wave packet
of a reference atom may be used to extract the HOMO from high
order harmonic spectra, may not be generalized to other molecules
such as CO$_2$. For these molecules, the molecular properties
already enter the recolliding electron wave packet so  the
spectral amplitude of the molecular REWP shows a large difference
from the reference atom and a strong orientation dependence in
some energy regions due to the interference between the atomic
cores of the molecule.

This work has been supported by NNSF Grant No. 10725521, CAEP
Foundation Project No. 2006Z0202, 973 Research Project No.
2006CB806000, and by National Fundamental Research Programme of
China under Grant Nos. 2006CB921400 and 2007CB814800.

\end{document}